\documentclass[preprint,showpacs,preprintnumbers,
amsmath,amssymb]{revtex4-1}

\usepackage{graphicx}
\usepackage{dcolumn}
\usepackage{bm,morefloats,color}

\usepackage{amssymb,bm}

\def\Eq#1{\begin{equation} #1 \end{equation}}
\def\Eqr#1{\begin{eqnarray} #1 \end{eqnarray}}
\def\Eqrsubl#1#2{\begin{subequations}\label{#1}\Eqr{#2}\end{subequations}}

\newcommand{\nn}{\nonumber}
\newcommand{\pd}{\partial}

\newcommand{\bea}{\begin{eqnarray}}
\newcommand{\eea}{\end{eqnarray}}

\def\Xsp{{\rm X}}

\def\Ysp{{\rm Y}}

\def\X5sp{{\rm X}_5}
\def\Y3sp{{\rm Y}_3}
\def\Z3sp{{\rm Z}_3}

\def\lap{{\triangle}}

\begin{document}

\title{
Dynamical branes on expanding orbifold and 
complex projective space
}

\author{Muneto Nitta}%
\affiliation{%
Department of Physics, and Research and Education 
Center for Natural Sciences, Keio University, 
Hiyoshi 4-1-1, Yokohama, Kanagawa 223-8521, Japan}%
\affiliation{
International Institute for Sustainability with Knotted Chiral Meta Matter(SKCM$^2$), Hiroshima University, 1-3-2 Kagamiyama, Higashi-Hiroshima, Hiroshima 739-8511, Japan}

\author{Kunihito Uzawa}
\affiliation{%
Department of Physics,
School of Science and Technology,
Kwansei Gakuin University, Sanda, Hyogo 669-1337, Japan
\\
and
\\
Research and Education 
Center for Natural Sciences, Keio University, 
Hiyoshi 4-1-1, Yokohama, Kanagawa 223-8521, Japan
}%

\date{\today}

\begin{abstract} 
We construct some new dynamical $p$-brane solutions 
to gravity theories on curved backgrounds. 
We discuss the relations 
between dynamical branes, a new time-dependent solution on 
complex projective space $\mathbb{C}{\rm P}^n$ and the static 
$p$-branes on the orbifold ${\mathbb C}^n/{\mathbb Z}_n$.
\end{abstract}

\pacs{
04.65.+e, 11.25.-w, 11.27.+d}

\maketitle


\section{Introduction}
\label{sec:introduction}

D$p$-branes or more generally $p$-branes are 
$(p+1)$-dimensional non-perturbative solitonic 
objects appearing in string theory and 
supergravity \cite{Townsend:1995gp, Argurio:1998cp, Duff:1990xz}.
Such brane configurations were applied in the contexts of brane world models 
and brane cosmology 
\cite{Gibbons:2005rt, 
Binetruy:2007tu, 
Maeda:2009zi, 
Minamitsuji:2010fp, Maeda:2010aj, Minamitsuji:2010kb, 
Minamitsuji:2010uz, 
Minamitsuji:2011jt, 
Maeda:2012xb, 
Minamitsuji:2012if, Uzawa:2013koa, 
Uzawa:2013msa, 
Uzawa:2014kka, Uzawa:2014dra, Maeda:2015joa, 
Maeda:2016lqw, Maeda:2017caf}.
In particular, in application to cosmology such as inflationary universe,   
branes cannot be static anymore, 
and dynamics of branes is essential  
\cite{Binetruy:2007tu, Maeda:2009zi, Minamitsuji:2010fp, 
Minamitsuji:2010kb, Minamitsuji:2010uz, Minamitsuji:2011jt, 
Maeda:2012xb, Minamitsuji:2012if, Uzawa:2013msa, Uzawa:2014kka}.
In black hole physics, dynamical branes are essential for 
black hole dynamics such as their collision 
\cite{Gibbons:2005rt, Maeda:2010aj, Maeda:2012xb, 
Uzawa:2014dra, Maeda:2016lqw}.
The dynamical $p$-brane solutions in a higher-dimensional 
gravity theory were studied in Refs.~\cite{Gibbons:2005rt, 
Binetruy:2007tu, 
Maeda:2009zi, 
Minamitsuji:2010fp, Maeda:2010aj, Minamitsuji:2010kb, 
Minamitsuji:2010uz, 
Minamitsuji:2011jt, 
Maeda:2012xb, 
Minamitsuji:2012if, Uzawa:2013koa, 
Uzawa:2013msa, 
Uzawa:2014kka, Uzawa:2014dra, Maeda:2015joa, 
Maeda:2016lqw, Maeda:2017caf} 
and have been widely discussed ever since. 
However, some aspects of the physical properties, 
such as having quadratic order of time dependence 
 and its dynamics in the context of 
string theory, have remained unclear. 
The motivation of this work is to improve this situation. For
this purpose, it is first necessary to construct a  
dynamical solutions depending on the time as well as space
coordinates.

There are also interesting works recently to find the 
dynamical $p$-brane solutions giving 
the dynamics of supersymmetry breaking \cite{Maeda:2017caf} and 
the issues of spacetime singularity such as 
cosmic censorship conjecture \cite{Maeda:2015joa}. 
Since some of dynamical solutions preserve supersymmetry, 
we can find the relation deeply between the expansion 
of universe and breaking of supersymmetry \cite{Maeda:2017caf}. 
The dynamical branes have been found by 
classical solutions of supergravities which are the 
low-energy effective theories of superstring theories or 
eleven-dimensional supergravity 
\cite{Maeda:2009zi}. 
Since the dynamical $p$-brane are extension to 
static $p$-brane in string theory which have been 
objects of intensive research, these objects have been treated as  
dynamical objects in general relativity 
as well as string theories. 
The dynamical $p$-brane solutions give interesting results 
and important descriptions of their dynamics 
in supergravities. 

In this paper, we find new dynamical $p$-brane solutions, 
which are classified into the two classes.
In the first, we promote 
$p$-brane solutions on the orbifolds ${\mathbb C}^n/{\mathbb Z}_n$ 
\cite{Nitta:2020pmp, Nitta:2020pzo} 
to dynamical ones. 
In this case, the orbifolds expand in time. 
The second is dynamical 
solution on the complex projective space ${\mathbb C}{\rm P}^n$. 

This paper is organized as follows. 
The section \ref{sec:n} gives a brief introduction to 
dynamical $p$-brane in gravity theory. 
The ansatz of fields and the various field equations 
are then discussed. Section \ref{sec:o} is presented 
for constructing dynamical $p$-brane solutions on the orbifold. 
The setup is a simple extension of the static $p$-brane 
system. Section \ref{sec:o} is devoted to constructing 
a new dynamical solutions carrying one antisymmetric 
tensor field charge. We discuss extremal (black) $p$-branes and 
dynamical solutions. 
When the space that gauge potential does not extend is   
non-Ricci flat, the function in the metric is no longer linear
in time like dynamical $p$-brane system 
but quadratic in it. We will show it in section \ref{sec:CP}. 
Although solutions we find in section \ref{sec:CP} do not describe 
a $p$-brane, it will allow us to obtain dynamical 
solutions in $D$-dimensional gravity theory. 
Finally, we conclude in Sec.\ref{sec:cr}.

\section{Charged extremal and dynamical black $p$-branes}
\label{sec:n} 
We briefly summarize the results for 
$(p+2)$-form field strength in the $D$-dimensional theory. 
We consider a gauge field strength $F_{(p+2)}$ in the action 
\cite{Nitta:2020pmp}
\Eq{
S=\frac{1}{2\kappa^2}\int d^Dx \sqrt{-g}
\left[R-\frac{1}{2\left(p+2\right)!}F_{(p+2)}^2\right],
   \label{n:action:Eq}
}
where $\kappa^2$ is the $D$-dimensional 
gravitational constant and $R$ denotes the Ricci scalar with 
respect to the $D$-dimensional metric $g_{MN}$. 
The field equations are given by
\Eqrsubl{n:fields:Eq}{
&&R_{MN}=\frac{1}{2\cdot \left(p+2\right)!}
\left[\left(p+2\right)F_{MA_1\cdots A_{p+1}}
{F_{N}}^{A_1\cdots A_{p+1}}-g_{MN}F_{(p+2)}^2\right],
  \label{n:Einstein:Eq}\\
&&d\left(\ast F_{(p+2)}\right)=0\,.
  \label{n:F:Eq}
}

We review the properties of the $p$-brane 
to simplify the field equations. 
The $p$-brane has $p$ spacelike directions 
which are longitudinal to the $p$-brane. It contains  
also $(D-p-1)$ spacelike directions that 
are characterized by transverse to the $p$-brane. 

The longitudinal spacetime to the $p$-brane thus gives 
the timelike direction. 
We will consider a single dynamical $p$-brane solutions 
with a single charge. 
The dynamical $p$-brane 
do not have translational invariant 
with respect to the longitudinal spacetime to the $p$-brane. 
Since they are localized at a point in the transverse space 
to the $p$-brane, there are also no translational invariance. 
We suppose spherical symmetry in
the $(D-p-1)$-dimensional transverse space for the 
dynamical $p$-brane without any angular momentum. 

We take a single $p$-brane ansatz for $D$-dimensional metric
\Eq{
ds^2=h^a(x,~y)\,q_{\mu\nu}\left(x\right)dx^\mu dx^\nu
+h^b(x,~y)\,u_{ij}\left(y\right)dy^i dy^j\,, 
    \label{n:metric:Eq}
}
where $q_{\mu\nu}\left(x\right)$ is a 
$(p+1)$-dimensional metric which 
depends only on the coordinates $t$\,, $x^{\alpha}$ with 
$\alpha$ being the spatial coordinates, and $u_{ij}
\left(y\right)$ is the $(D-p-1)$-dimensional metric 
which depends only on the coordinates $y^i$.  
The coordinates of $D$-dimensional spacetime are divided by 
two sets, $x^M=(x^\mu, y^i)$, 
with $\mu=0, \cdots, p$ and $i=1,\cdots, D-p-1$.
Here, the $y^i$'s denote the coordinates of
 the transverse space. 
We divide again the coordinates $x^\mu$ into two parts; 
the time coordinate $t$ and the spatial coordinates 
$x^\alpha~(\alpha=1, \cdots, p)$\,, where the 
$x^\alpha$'s 
span the directions longitudinal to the brane. 
We choose the time-like direction $x^0=t$ and 
assume that the metric depends on only $t$ and $y^i$, 
but also on $x^\alpha$\,. The metric form (\ref{n:metric:Eq}) 
is a straightforward generalization of the case of a static 
$p$-brane system \cite{Gibbons:2005rt, Binetruy:2007tu, 
Maeda:2009zi}.

The parameters $a$ and $b$ in the dynamical brane system 
are given by 
\Eq{
a=-\frac{D-p-3}{D-2}\,,~~~~~b=\frac{p+1}{D-2}\,,
   \label{n:ab:Eq}
}
while the gauge field strength $F_{(p+2)}$ is also assumed to be
\Eq{
F_{(p+2)}=d\left(h^{-1}\right)\wedge dt\wedge dx^1\wedge \cdots 
\wedge dx^p\,.
  \label{n:gauge:Eq}
}
Under our ansatz, the Einstein equations become
\Eqrsubl{n:cEinstein:Eq}{
&&R_{\mu\nu}(\Xsp)-h^{-1}D_{\mu}D_{\nu}h
-\frac{a}{2}h^{-1}q_{\mu\nu}
\left(\lap_\Xsp h+h^{-1}\lap_\Ysp h\right)=0\,,
  \label{n:cEinstein-mn:Eq}\\
&&R_{ij}(\Ysp)-\frac{b}{2}
u_{ij}\left(\lap_\Xsp h+h^{-1}\lap_\Ysp h\right)=0\,, 
  \label{n:cEinstein-ij:Eq}\\
&&\pd_\mu\pd_ih=0\,,
   \label{n:cEinstein-mi:Eq}
}
where $D_\mu$ denotes the covariant derivative with 
respective to the metric $q_{\mu\nu}$, $\lap_\Xsp$ and 
$\lap_\Ysp$ are the Laplace operators on the 
$(p+1)$-dimensional world-volume spacetime X and 
$(D-p-1)$-dimensional space Y 
spaces, and $R_{\mu\nu}(\Xsp)$ and $R_{ij}(\Ysp)$ are 
the Ricci tensors of the metrics $q_{\mu\nu}$ and $u_{ij}$, 
respectively. From Eq.~(\ref{n:cEinstein-mi:Eq}), 
the function $h(x, y)$ have to be in the form
\Eq{
h(x, y)=h_0(x)+h_1(y). 
}
The other components of the Einstein equations 
(\ref{n:cEinstein-mn:Eq}) and (\ref{n:cEinstein-ij:Eq}) can be 
rewritten as
\Eqrsubl{n:cEinstein2:Eq}{
&&R_{\mu\nu}(\Xsp)-h^{-1}D_{\mu}D_{\nu}h_0
-\frac{a}{2}h^{-1}q_{\mu\nu}
\left(\lap_\Xsp h_0+h^{-1}\lap_\Ysp h_1\right)=0\,,
  \label{n:cEinstein2-mn:Eq}\\
&&R_{ij}(\Ysp)-\frac{b}{2}u_{ij}
\left(\lap_\Xsp h_0+h^{-1}\lap_\Ysp h_1\right)=0\,. 
  \label{n:cEinstein2-ij:Eq}
}
Next we consider the gauge field strength. From the assumption 
(\ref{n:gauge:Eq}), we find that the Bianchi identity is automatically 
satisfied while the equation of motion for the
gauge field (\ref{n:F:Eq}) becomes $\lap_\Ysp h=0$\,, 
and $\pd_\mu\pd_i h=0$\,. 

If $F_{(p+2)}\ne 0$, the function $h_1$ is non-trivial. 
The Einstein equations thus reduce to
\Eqrsubl{n:Einstein2:Eq}{
&&R_{\mu\nu}(\Xsp)=0\,,
   \label{n:E2-1:Eq}\\
&&R_{ij}(\Ysp)=\frac{1}{2}b\left(p+1\right)\lambda 
\,u_{ij}\,,
   \label{n:E2-2:Eq}\\
&&D_\mu D_\nu h_0=\lambda\,q_{\mu\nu}\,,
   \label{n:E2-3:Eq}
}
where $\lambda$ is a constant. We see that the space Y  
is not Ricci flat, but the Einstein space such as 
$\mathbb{C}$P${}^{n}$ if $\lambda\ne 0$, 
and the function $h$ can be more non-trivial.  

Let us consider the case 
\Eq{
q_{\mu\nu}=\eta_{\mu\nu}\,,
}
where X is $(p+1)$-dimensional Minkowski spacetime. 
If $D_\mu h_0\ne 0$ and $\left(D_\mu h_0\right)
\left(D^\mu h_0\right)\ne 0$\,, 
the solution for $h_0$ is given by
\Eq{
h_0(x)=\frac{\lambda}{2}x_\mu x^\mu 
+\bar{a}_\mu x^\mu +\bar{a}\,.
  \label{h-sol}
}
Here we have introduced constants 
$\bar{a}_\mu$ and $\bar{a}$ satisfying the condition 
$\bar{a}_\mu\,\bar{a}^\mu \ne 0$\,. However, 
if $D_\mu h_0\ne 0$ and 
$\left(D_\mu h_0\right)\left(D^\mu h_0\right)=0$\,, 
there exists a solution only when $\lambda=0$\,.

Before concluding this section, we should comment the ansatz 
for fields (\ref{n:Einstein2:Eq}). The simplification to the 
field equations (\ref{n:fields:Eq}) strongly depends on 
choosing parameters for the metric. With this choice, the 
metric can be written by the function $h(x, y)$ multiplying 
a flat metric for the $(p+1)$-dimensional longitudinal spacetime. 
Note that the function $h(x, y)$ depends on 
$(D-p-1)$-dimensional transverse 
space to the $p$-brane as well as the $(p+1)$-dimensional 
longitudinal coordinates. 
Hence, the $p$-brane is fully characterized by time. 
Moreover, in the context of cosmology, dynamical $p$-brane 
 is most of the time related to the fact that the solutions 
describes an expansion of universe.

%
\section{Dynamical $p$-brane on orbifold}
\label{sec:o}
We now construct the solution of dynamical $p$-brane on orbifold 
explicitly. This case is interesting because 
field equations are analytically solved (\ref{n:Einstein2:Eq}). 

The Einstein equations (\ref{n:E2-2:Eq}) 
can be solved when we 
start with a $\mathbb{C}$P${}^{D-p-3}$ metric 
in $(D-p-1)$ dimensions, namely \cite{Hoxha:2000jf}:
\Eq{
u_{ij}(y)dy^idy^j=
dr^2+r^2\left[\left\{d\rho+2n\,\sin^2\xi_{n-1}
\left(d\psi_{n-1}+\frac{1}{2(n-1)}\omega_{n-2}\right)
\right\}^2+ds^2_{\mathbb{C}{\rm P}^{n-1}}\right],
  \label{o:metric-y:Eq}
}
where $r$ is a radial coordinate, 
$\rho$ is a coordinate of $S^1$, 
$\xi_{n-1}$ and $\psi_{n-1}$ are coordinates of 
the $\mathbb{C}{\rm P}^{n-1}$ space 
with the ranges $0\le\xi_{n-1}\le \pi/2$\,,~
$0\le \psi_{n-1}\le 2\pi$\,,  
$\omega_{n-1}$ and 
$ds^2_{\mathbb{C}{\rm P}^{n-1}}$ state 
a one-form and a metric on the  
${\mathbb{C}}{\rm P}^{n-1}$ space, 
recursively defined as
\cite{Dehghani:2005zm, Dehghani:2006aa, Tatsuoka:2011tx}
\Eqr{
ds^2_{\mathbb{C}{\rm P}^{n-1}}&=&
2n\left[d\xi_{n-1}^2+
\sin^2\xi_{n-1}\,\cos^2\xi_{n-1}
\left\{d\psi_{n-1}+\frac{1}{2(n-1)}
\omega_{n-2}\right\}^2\right.\nn\\
&&\left.+\frac{1}{2(n-1)}\sin^2\xi_{n-1}\,
ds^2_{\mathbb{C}{\rm P}^{n-2}}
\right],
} 
and 
\Eqrsubl{cpn:cp1:Eq}{
\omega_{n-2}&=&2\left(n-1\right)\sin^2\xi_{n-2}
\left[d\psi_{n-2}+\frac{1}
{2\left(n-2\right)}\omega_{n-3}\right],\\
ds^2_{\mathbb{C}{\rm P}^1}&=&4\left(
d\xi_1^2+\sin^2\xi_1\,\cos^2\xi_1\,d\psi_1^2\right)\,,\\
\omega_1&=&4\sin^2\xi_1\,d\psi_1\,.
}
Here $(r, \rho)$ describes a complex line,
and $\rho$ together with ${\mathbb C}{\rm P}^n$ 
denote a $(2n-1)$-dimensional sphere 
$S^{2n-1}/{\mathbb Z}_n = S^{D-p-2}/{\mathbb Z}_n $\,, 
which is actually an event horizon.

One remark is in order before we continue. 
In Eq.~(\ref{o:metric-y:Eq}), we assume 
$R(\Ysp)=0$\,, which is constructed from the 
metric $u_{ij}(y)$\,. 
Such an assumption would give $\lambda=0$ 
in the Einstein equations (\ref{n:E2-2:Eq}). 

We impose on the condition $h_1=h_1(r)$ in the field equations. 
If we introduce 
the dependence of radial coordinate $r$
for the function $h_1$\,, we have reduced 
the problem to the equation
\Eq{
\lap_\Ysp h_1=\frac{1}{r^{D-p-2}}\frac{d}{dr}
\left(r^{D-p-2}\frac{d}{dr}h_1\right)=0\,.
   \label{o:h1:Eq}
}
This is solved to give for $D-p-3\ne 0$
\Eq{
h_1(r)=b_1+\frac{b_2}{r^{D-p-3}}\,.
     \label{o:h1-2:Eq}
}
The constant parameters in Eq.~(\ref{o:h1:Eq}) 
are determined so that 
the solution is not singular for $r>0$ with $h_0=0$\,. 
The gauge field strength is asymptotically 
vanishing according to the limit 
$r\rightarrow\infty$ in the function $h_1(r)$. 
We have assumed $D-p-3>0$ in Eqs.~(\ref{o:h1:Eq}) and 
(\ref{o:h1-2:Eq}) giving zeroth of the gauge field strength 
asymptotically and a Kasner spacetime at infinity. 
We will discuss them more detail later. 
One can show that the solution (\ref{o:h1-2:Eq}), 
when $D-p-3\ne 0$ is replaced for $D-p-3=0$\,, 
is a direct consequence of Eq.~(\ref{o:h1:Eq}). 
The solution is then shown to be 
\Eq{
h_1(r)=b_3+b_4\ln r\,,
}
where 
the function 
$h_1$ diverges both at $r\rightarrow\infty$ 
and $r\rightarrow 0$\,. 
Since there is no regular spacetime region near $p$-brane, 
such solutions are not physically relevant. 
We will here consider the case $D-p-3>0$ in the following.
We have constructed dynamical solutions depending on
parameters, $\bar{a}_\mu$\,, $\hat{a}$\,, and 
$b_2$\,. The solutions are characterized by the function 
which is harmonic in $(D-p-3)$-dimensional space:
\Eq{
h(x, r)=\bar{a}_\mu x^\mu+\hat{a}
+\frac{b_2}{r^{D-p-3}}\,,
   \label{n:th:Eq}
}
where $\hat{a}$ is defined by 
$\hat{a}=\tilde{a}+b_1$\,. 

The surfaces of constant $t$ are spacelike everywhere. 
The geometry resembles 
the infinite throat familiar from the asymptotically flat extremal
Reissner-Nordstr\"{o}m solution near $r=0$\,. 
This can be expressed by the spatial metric in
spherical coordinates centered at $r=0$. 
Near the origin of these coordinates, this metric becomes
\Eqr{
ds^2\approx\left(\frac{b_2}{r^{D-p-3}}\right)^a
\,\eta_{\mu\nu}
+\left(\frac{b_2}{r^{D-p-3}}\right)^b\,r^2
\left(\frac{dr^2}{r^2}+d\Omega^2\right)\,,
}
which is the metric for a warped cylinder of infinite spatial 
extent having cross sectional area. 
Here the metric $d\Omega^2$ takes the form of 
\Eq{
d\Omega^2=\left[d\rho+2n\,\sin^2\xi_{n-1}
\left(d\psi_{n-1}+\frac{1}{2(n-1)}\omega_{n-2}\right)
\right]^2+ds^2_{\mathbb{C}{\rm P}^{n-1}}\,.
}

If we set $D-p-3=-1$ and $\bar{a}_\alpha=0~(\alpha=1, 2, \cdots, p)$, 
then we have $h(t,~r)=\bar{a}_0\,t+\hat{a}+b_2\,r$\,. 
Hence any points on the branes are regular, and time dependent.
When we take the limit of $h(t,~r)\rightarrow 0$ (or finite) as 
$r\rightarrow \infty$ for $D-p-3>1$ (or $r$ is finite for 
$D-p-3=1$), the spacetime turns out to be time dependent. 
To see its dynamical behaviour, we introduce a new time coordinate
\Eqr{
\tau=\tau_0\left(\bar{a}_0\,t\right)^{\frac{a}{2}+1}\,,
}
where $\tau=\frac{2}{\bar{a}_0(a+2)}$\,.
The asymptotic dynamical solution is rewritten as
\Eqr{
ds^2=-d\tau^2+\left(\frac{\tau}{\tau_0}\right)^
{a\left(\frac{a}{2}+1\right)^{-1}}
\sum_\alpha\left(dx^\alpha\right)^2
+\left(\frac{\tau}{\tau_0}\right)^
{b\left(\frac{a}{2}+1\right)^{-1}}u_{ij}dy^idy^j\,.
} 
Hence, we find a Kasner-like expansion:
\Eqrsubl{Kasner}{
&&\frac{a}{2}\left(\frac{a}{2}+1\right)^{-1}\,p
+\frac{b}{2}\left(\frac{a}{2}+1\right)^{-1}
(D-p-1)=1\,,
  \label{Kasner1}\\
&&\frac{a^2}{4}\left(\frac{a}{2}+1\right)^{-2}\,p
+\frac{b^2}{4}\left(\frac{a}{2}+1\right)^{-2}(D-p-1)=1\,.
  \label{Kasner2}
}
Eq.~(\ref{Kasner1}) is always satisfied for any dynamical 
$p$-brane configuration while Eq.~(\ref{Kasner2})
is true only for M-theory or D3-brane system 
because there is no or trivial dilaton 
in the background \cite{Maeda:2009zi}.
The dynamics of brane is also correct when 
we fix the position in the transverse space 
to the $p$-brane, even if the
metric is locally inhomogeneous in the bulk space.

The curvature of Eqs.~(\ref{n:metric:Eq}) and 
(\ref{n:th:Eq}) can be singular at zeros of the 
metric function $h$\,. This can be seen from the square of 
the $(p+2)$-form field strength,
\Eq{
F_{(p+2)}^2\equiv F_{A_1\cdots A_{p+2}}
F^{A_1\cdots A_{p+2}}
=-h^\alpha\,\left(\pd_r h\right)^2\,,
}
where $\alpha=-4+\frac{(D-p-4)(p+1)}{D-2}<0$\,. 
If the function $h=0$ and $\pd_rh$ does not vanish 
like $h^2$ or faster, then $F_{(p+2)}^2$ diverges 
and the curvature is singular.

According to these elements, we can find a 
behaviour of how the background geometry 
develops in time. 
Since the function $h$ is positive everywhere 
for $\bar{a}_\mu x^\mu+\hat{a}>0$\,, the spatial surfaces are 
not singular. They are asymptotically time dependent
spacetime and have the cylindrical form of an 
infinite throat near $r=0$\,. The spatial metric 
is not singular and the cylindrical form 
everywhere. When $\bar{a}_\mu x^\mu+\hat{a}$ is slightly increased, 
a singularity appears near $r=\infty$\,. As 
$\bar{a}_\mu x^\mu+\hat{a}$ increases further, the singularity 
cuts off more and more of the cylinder.

\section{Dynamical solution on $\mathbb{C}$P${}^{n}$ space}
\label{sec:CP}

In this section, we present the dynamical solution on 
the $\mathbb{C}$P${}^{n}$ space which happens when the
Einstein equations become Eq.~(\ref{n:Einstein2:Eq}).
As seen from the Einstein equations, the
internal space Y is not necessarily Ricci flat, and the
function $h_0$ is no longer linear in the coordinates 
$x^\mu$ but quadratic in them. 

%
\subsection{Dynamical solution on $\mathbb{C}$P${}^{1}$ space}

First, we consider the case in which Y is a simple 
$\mathbb{C}$P${}^{1}$ space 
\Eq{
ds^2_{\mathbb{C}{\rm P}^1}=\left(1+\tilde{r}^2\right)^{-2}
\left(d\tilde{r}^2+\tilde{r}^2d\tilde{\theta}^2\right)\,.
}
Note that $\mathbb{C}$P${}^{1}$ space 
can be expressed by 
the Fubini-Study metric because of a diffeomorphism 
$\mathbb{C}$P${}^{1}\cong S^2$\,.
Let $h_1(\tilde{r},\,\tilde{\theta})$ be a function on Y of the form
\Eq{
h_1(\tilde{r}\,, \tilde{\theta})
=\tilde{H}(\tilde{r})+\tilde{K}(\tilde{\theta})\,.
}
Then, the equation $\lap_\Ysp h_1=0$ gives 
\Eq{
\pd_{\tilde{r}}\left(\tilde{r}\,\pd_{\tilde{r}}\tilde{H}\right)
+\frac{1}{\tilde{r}}\,\pd_{\tilde{\theta}}^2\tilde{K}=0\,.
}
If we assume that functions $\tilde{H}(\tilde{r})$ and 
$\tilde{K}(\tilde{\theta})$ obey 
\Eq{
\pd_{\tilde{r}}\left(\tilde{r}\,\pd_{\tilde{r}}\tilde{H}\right)
=0\,,~~~~~
\frac{1}{\tilde{r}}\,\pd_{\tilde{\theta}}^2\tilde{K}
=0\,,
}
we find 
\Eq{
h_1(\tilde{r},\,\tilde{\theta})=
\tilde{c}_1\ln \tilde{r}+\tilde{c}_2\,\tilde{\theta}
+\tilde{c}_3\,.
  \label{cp1:h1:Eq}
}
Here $\tilde{c}_i~(i=1\,,\cdots\,,3)$
are constants.

The metric we found as the solution (\ref{h-sol}) and 
(\ref{cp1:h1:Eq}) is not of the product-type. 
The existence of a nontrivial gauge field strength forces
the function $h(x, y)$ to be a linear combination of a function of
$x^\mu$ and a function of $y^i$, which is not 
the conventional assumption. The function in Eq.~(\ref{h-sol}) 
implies that we cannot drop the dependence on the world 
volume coordinate for a non-vanishing Ricci scalar $R(\Ysp)$\,. 
This solution gives the inhomogeneous universe due to 
the function $h_1$ when we regard the bulk transverse space 
as four-dimensional space.

\subsection{Dynamical solution on $\mathbb{C}$P${}^{2}$ space}
Next, we discuss 
solution on $\mathbb{C}$P${}^{2}$ space, whose metric is given by 
\cite{Gibbons:1978zy}:
\Eq{
ds^2_{\mathbb{C}{\rm P}^2}=\left(1+\bar{\rho}^2\right)^{-2}
d\bar{\rho}^2
+\frac{\bar{\rho}^2}{4}\left(1+\bar{\rho}^2\right)^{-2}\left(
d\psi+\cos\theta d\phi\right)^2
+\frac{\bar{\rho}^2}{4}\left(1+\bar{\rho}^2\right)^{-1}\left(
d\theta^2+\sin^2\theta d\phi^2\right)\,.
}

If we set 
\Eq{
h_1(\bar{\rho},\,\theta)=\bar{H}(\bar{\rho})+\bar{K}(\theta)\,,
}
the equation $\lap_\Ysp h_1=0$ yields
\Eq{
\frac{\left(1+\bar{\rho}^2\right)^3}{\bar{\rho}^3}
\pd_{\bar{\rho}}\left(\frac{\bar{\rho}^3}{1+\bar{\rho}^2}
\pd_{\bar{\rho}} \bar{H}\right)+
\frac{1}{\sin\theta}
\pd_\theta\left[
\frac{4\left(1+\bar{\rho}^2\right)}{\bar{\rho}^2}
\sin\theta\,\pd_\theta \bar{K}\right]=0\,.
}
For example, we require that the functions $\bar{H}(\bar{\rho})$ and 
$\bar{K}(\theta)$ satisfy 
\Eq{
\frac{\left(1+\bar{\rho}^2\right)^3}{\bar{\rho}^3}
\pd_{\bar{\rho}}\left(\frac{\bar{\rho}^3}{1+\bar{\rho}^2}
\pd_{\bar{\rho}} \bar{H}\right)
=0\,,~~~~~~
\frac{1}{\sin\theta}
\pd_\theta\left[
\frac{4\left(1+\bar{\rho}^2\right)}{\bar{\rho}^2}
\sin\theta\,\pd_\theta \bar{K}\right]
=0\,.
}
The solution to these equations is
\Eqr{
\bar{H}(\bar{\rho})=\bar{c}_1\left(-\frac{1}{2\bar{\rho}^2}
+\ln \bar{\rho}\right)
+\bar{c}_2\,,~~~~~\bar{K}(\theta)=
\bar{c}_3\ln\tan\frac{\theta}{2}
+\bar{c}_4\,,
\label{HK}
}
where $\bar{c}_i~(i=1\,,\cdots\,,4)$ are constants. 

The scale factor of universe again includes the 
inhomogeneity due to functions $h_0$ and $h_1$. 
We live in the three-dimensional space after 
compactifying the $(p-3)$-dimensional space.  
In this case, since we fix our universe at some
position in the $\mathbb{C}$P${}^{2}$ space, 
the line element is given by
\Eqr{
ds^2=\left[\frac{\lambda}{2}\left(-t^2+
x^\alpha x_\alpha\right)\right]^a
\left(-dt^2+d\bar{r}^2+\bar{r}^2\,d\Omega^2_{(p-1)}\right)
+\left[\frac{\lambda}{2}\left(-t^2+
x^\alpha x_\alpha\right)\right]^b\,ds^2(\Ysp)\,,
   \label{cp2}
}
where we set $\bar{a}_\mu=0$\,, $\hat{a}=0$\,, 
$d\Omega^2_{(p-1)}$ denotes the metric of $(p-1)$-dimensional 
sphere S${}^{p-1}$\,, and 
\Eqrsubl{cp2-2}{
&&\eta_{\mu\nu}dx^\mu dx^\nu=
-dt^2+\delta_{\alpha\beta}dx^\alpha dx^\beta
=-dt^2+d\bar{r}^2+\bar{r}^2\,d\Omega^2_{(p-1)}\,,\\
&&ds^2(\Ysp)=dr^2+r^2\left[\left\{d\rho+6\sin^2\xi_2
\left(d\psi_2+\frac{1}{4}\omega_1\right)
\right\}^2
\right].
}
Although it looks inhomogeneous at first glance, 
in terms of the coordinate transformation for 
$\lambda>0$\,, 
\Eqr{
t=\sqrt{\frac{2}{\lambda}}\,T\,\sinh \bar{R}\,,~~~~
\bar{r}=\sqrt{\frac{2}{\lambda}}\,T\,\cosh \bar{R}\,,
}
we can rewrite the metric (\ref{cp2}) as
\Eqr{
ds^2&=&\frac{2}{\lambda}T^{2a}
\left[-dT^2+T^2\left\{
d\bar{R}^2+\bar{R}^2\,d\Omega^2_{(p-1)}
+d\bar{s}^2(\Ysp)\right\}\right]\nn\\
&=&\frac{2}{\lambda}\left[
-d\bar{T}^2+\left(\frac{p+1}{D-2}\right)^2\,
\bar{T}^2\,\left\{d\bar{R}^2+\bar{R}^2\,d\Omega^2_{(p-1)}
+d\bar{s}^2(\Ysp)\right\}
\right],
  \label{cp3}
}
where we have defined 
\Eqr{
\bar{T}=\left(\frac{D-2}{p+1}\right)\,
T^{(p+1)/(D-2)}\,,~~~~
d\bar{s}^2(\Ysp)=\frac{\lambda}{2}
\,ds^2(\Ysp)\,.
}
One can note that the metric (\ref{cp3}) represents an 
isotropic and homogeneous spacetime.  
The scale factor of universe is thus proportional 
to the function $\bar{T}$ of the cosmic time, 
which is known as the Milne universe.  

If $\lambda<0$, we should use the coordinate transformation:
\Eqr{
t=\sqrt{-\frac{2}{\lambda}}\,\bar{R}\,\cosh T\,,~~~~
\bar{r}=\sqrt{-\frac{2}{\lambda}}\,\bar{R}\,\sinh T\,.
}
Then we have 
\Eqr{
ds^2&=&\frac{2}{|\lambda|}\bar{R}^{2a}
\left[d\bar{R}^2+\bar{R}^2\left\{-dT^2
+\bar{R}^2\,d\Omega^2_{(p-1)}
+d\hat{s}^2(\Ysp)\right\}\right]\nn\\
&=&\frac{2}{|\lambda|}\left[
d\hat{R}^2+\left(\frac{p+1}{D-2}\right)^2\,
\hat{R}^2\,\left\{-dT^2+\hat{R}^2\,d\Omega^2_{(p-1)}
+d\hat{s}^2(\Ysp)\right\}
\right],
  \label{cp4}
}
where $\hat{R}$ and $d\hat{s}^2(\Ysp)$ are given by
\Eqr{
\hat{R}=\left(\frac{D-2}{p+1}\right)\,
\bar{R}^{(p+1)/(D-2)}\,,~~~~
d\hat{s}^2(\Ysp)=\frac{|\lambda|}{2}
\,ds^2(\Ysp)\,.
}
The metric (\ref{cp4}) describes a conformally flat and 
inhomogeneous spacetime, but it is different from a Milne universe.

\section{Conclusion and remarks}
\label{sec:cr}
We have found two new dynamical $p$-brane solutions.
The first is $p$-brane solutions on the orbifolds 
${\mathbb C}^n/{\mathbb Z}_n$ \cite{Nitta:2020pmp, Nitta:2020pzo} 
 expanding in time. 
The second is dynamical solutions on the complex projective 
space ${\mathbb C}{\rm P}^n$. 

Our new solutions have been obtained by replacing a 
constant $c$ in the function $h=c+h_1$ of a 
static solution with a quadratic function of 
the coordinates $x^\mu$. 
We have obtained dynamical $p$-brane solutions on the orbifold 
whose spacetime metric depends on the coordinates of 
both the worldvolume and the space transverse to the $p$-brane. 
The field equations normally indicate that 
dynamical solutions can be found while 
two functions in the metric depends on both the 
time and overall transverse space coordinates. 
We have constructed a solution explicitly
 in the case of $\lambda\ne 0$ beyond the examples considered 
in the previous works. 
The ansatz for fields to solve the field equations 
have been chosen by the extension to the 
static solution or the supersymmetric static 
$p$-brane solution, which is the extremal case. 
We have proceeded the construction further with respect to
the $D$-dimensional action (\ref{n:action:Eq}) and 
considered a time-dependent gauge field strength 
in the background. Since the field equation with our ansatz 
of fields allows the time-dependent solution, 
the supergravity theories, for instance, 
realize the dynamical $p$-brane at the classical level.
We could present dynamical solution explicitly in 
Eqs.~(\ref{cp1:h1:Eq}) and (\ref{HK}). 
We note that the no-force condition for the dynamical 
$p$-brane on the orbifold is the same as dynamical branes 
which have been discussed in \cite{Uzawa:2014kka}.

Constructing dynamical $p$-brane solutions on the orbifold are 
most interesting issues of the string cosmology because 
the evolution of universe is 
derived from brane configurations.
We then find cosmological models from those
solutions by smearing some dimensions. 
We have the cosmological solutions with a power-law expansion. 
However, the solutions of Einstein equations cannot give a 
realistic expansion law. 
Although our solution gives the dynamics of the 
various branes in $D$-dimensions, we have to specify 
the compactification to construct the 
four-dimensional cosmology. 
The time-dependent solution we have obtained here 
would give a key to construct in more realistic 
cosmological models. 

\section*{Acknowledgments}

This work of M.N. is supported in part by 
Grant-in-Aid for Scientific Research, 
 JSPS KAKENHI (Grant Number JP22H01221) 
and the WPI program ``Sustainability with Knotted Chiral Meta Matter (SKCM$^2$)'' at Hiroshima University.
The work of K. U. is supported by Grants-in-Aid from the Scientific 
Research Fund of the Japan Society for the Promotion of 
Science, under Contract No. 16K05364 and by the Grant ``Fujyukai'' 
from Iwanami Shoten, Publishers. 





\begin{thebibliography}{99}

\bibitem{Townsend:1995gp}
P.~K.~Townsend,
``P-brane democracy,''
[arXiv:hep-th/9507048 [hep-th]]. 

\bibitem{Argurio:1998cp}
  R.~Argurio,
  ``Brane physics in M theory,''
  hep-th/9807171.

\bibitem{Duff:1990xz}
  M.~J.~Duff and K.~S.~Stelle,
  ``Multimembrane solutions of $D=11$ supergravity,''
  Phys.\ Lett.\ B {\bf 253} (1991) 113.

\bibitem{Gibbons:2005rt}
  G.~W.~Gibbons, H.~Lu and C.~N.~Pope,
  ``Brane worlds in collision,''
  Phys.\ Rev.\ Lett.\  {\bf 94} (2005) 131602
  [arXiv:hep-th/0501117].

\bibitem{Binetruy:2007tu}
  P.~Binetruy, M.~Sasaki and K.~Uzawa,
  ``Dynamical D4-D8 and D3-D7 branes in supergravity,''
  Phys.\ Rev.\  D {\bf 80} (2009) 026001
  [arXiv:0712.3615 [hep-th]].

\bibitem{Maeda:2009zi}
  K.~i.~Maeda, N.~Ohta and K.~Uzawa,
  ``Dynamics of intersecting brane systems -- Classification and their
  applications --,''
  JHEP {\bf 0906} (2009) 051
  [arXiv:0903.5483 [hep-th]]. 

\bibitem{Minamitsuji:2010fp}
M.~Minamitsuji, N.~Ohta and K.~Uzawa,
``Dynamical solutions in the 3-Form Field Background in the Nishino-Salam-Sezgin Model,''
Phys. Rev. D \textbf{81} (2010), 126005
[arXiv:1003.5967 [hep-th]].

\bibitem{Maeda:2010aj}
K.~i.~Maeda, M.~Minamitsuji, N.~Ohta and K.~Uzawa,
``Dynamical $p$-branes with a cosmological constant,''
Phys. Rev. D \textbf{82} (2010), 046007
[arXiv:1006.2306 [hep-th]].

\bibitem{Minamitsuji:2010kb}
M.~Minamitsuji, N.~Ohta and K.~Uzawa,
``Cosmological intersecting brane solutions,''
Phys. Rev. D \textbf{82} (2010), 086002
[arXiv:1007.1762 [hep-th]].

\bibitem{Minamitsuji:2010uz}
M.~Minamitsuji and K.~Uzawa,
``Cosmology in $p$-brane systems,''
Phys. Rev. D \textbf{83} (2011), 086002
[arXiv:1011.2376 [hep-th]].

\bibitem{Minamitsuji:2011jt}
M.~Minamitsuji and K.~Uzawa,
``Dynamics of partially localized brane systems,''
Phys. Rev. D \textbf{84} (2011), 126006
[arXiv:1109.1415 [hep-th]].

\bibitem{Maeda:2012xb}
K.~i.~Maeda and K.~Uzawa,
``Dynamical brane with angles: Collision of the universes,''
Phys. Rev. D \textbf{85} (2012), 086004
[arXiv:1201.3213 [hep-th]].

\bibitem{Minamitsuji:2012if}
M.~Minamitsuji and K.~Uzawa,
``Cosmological brane systems in warped spacetime,''
Phys. Rev. D \textbf{87} (2013) no.4, 046010
[arXiv:1207.4334 [hep-th]].

\bibitem{Uzawa:2013koa}
K.~Uzawa and K.~Yoshida,
``Dynamical Lifshitz-type solutions and aging phenomena,''
Phys. Rev. D \textbf{87} (2013) no.10, 106003
[arXiv:1302.5224 [hep-th]].

\bibitem{Uzawa:2013msa}
K.~Uzawa and K.~Yoshida,
``Dynamical F-strings intersecting D2-branes in type IIA supergravity,''
Phys. Rev. D \textbf{88} (2013), 066005
[arXiv:1307.3093 [hep-th]].

\bibitem{Uzawa:2014kka}
K.~Uzawa and K.~Yoshida,
``Probe brane dynamics on cosmological brane backgrounds,''
Phys. Lett. B \textbf{738} (2014), 493-498
[arXiv:1401.3664 [hep-th]].

\bibitem{Uzawa:2014dra}
K.~Uzawa,
``Colliding $p$-branes in the dynamical intersecting brane system,''
Phys. Rev. D \textbf{90} (2014) no.2, 025024
[arXiv:1407.7406 [hep-th]].

\bibitem{Maeda:2015joa}
K.~Maeda and K.~Uzawa,
``Violation of cosmic censorship in dynamical $p$-brane systems,''
Phys. Rev. D \textbf{93} (2016) no.4, 044003
[arXiv:1510.01496 [hep-th]].

\bibitem{Maeda:2016lqw}
K.~i.~Maeda and K.~Uzawa,
``Dynamical angled brane,''
Phys. Rev. D \textbf{94} (2016) no.12, 126016
[arXiv:1603.01948 [hep-th]].

\bibitem{Maeda:2017caf}
K.~Maeda and K.~Uzawa,
``Supersymmetry in a dynamical M-brane background,''
Phys. Rev. D \textbf{96} (2017) no.8, 084053
[arXiv:1705.09878 [hep-th]].

\bibitem{Nitta:2020pzo}
M.~Nitta and K.~Uzawa,
``Orbifold black holes,''
Eur. Phys. J. C \textbf{81} (2021) no.6, 513
[arXiv:2011.13316 [hep-th]].

\bibitem{Nitta:2020pmp}
M.~Nitta and K.~Uzawa,
``Fractional black $p$-branes on orbifold ${\mathbb C}^n/{\mathbb Z}_n$,''
JHEP \textbf{03} (2021), 018
[arXiv:2012.13285 [hep-th]].

\bibitem{Hoxha:2000jf}
P.~Hoxha, R.~R.~Martinez-Acosta and C.~N.~Pope,
``Kaluza-Klein consistency, Killing vectors, and Kahler spaces,''
Class. Quant. Grav. \textbf{17} (2000), 4207-4240
[arXiv:hep-th/0005172 [hep-th]].

\bibitem{Dehghani:2005zm}
  M.~H.~Dehghani and R.~B.~Mann,
  ``NUT-charged black holes in Gauss-Bonnet gravity,''
  Phys.\ Rev.\ D {\bf 72} (2005) 124006
  [hep-th/0510083].

\bibitem{Dehghani:2006aa}
  M.~H.~Dehghani and S.~H.~Hendi,
  ``Taub-NUT/bolt black holes in Gauss-Bonnet-Maxwell gravity,''
  Phys.\ Rev.\ D {\bf 73} (2006) 084021
  [hep-th/0602069].

\bibitem{Tatsuoka:2011tx}
T.~Tatsuoka, H.~Ishihara, M.~Kimura and K.~Matsuno,
``Extremal Charged Black Holes with a Twisted Extra Dimension,''
Phys. Rev. D \textbf{85} (2012), 044006
[arXiv:1110.6731 [hep-th]].

\bibitem{Gibbons:1978zy}
  G.~W.~Gibbons and C.~N.~Pope,
  ``CP${}^2$ As A Gravitational Instanton,''
  Commun.\ Math.\ Phys.\  {\bf 61} (1978) 239.

\end{thebibliography}
\end{document}